\documentclass[a4paper,11pt]{article}
\usepackage[left=3.17cm,right=3.17cm,top=2.54cm,
headheight=0.5cm,headsep=0.54cm,bottom=2.54cm,footskip=0.79cm
]{geometry}

\usepackage[all]{xy}
\usepackage{amsmath,amsfonts,amssymb,amsthm,epsfig,amscd,comment,latexsym,psfrag}
\usepackage{nicefrac,xspace,tikz}

\DeclareMathOperator{\res}{res}

\usetikzlibrary{arrows}
\usetikzlibrary{decorations.markings}
\def\Z{\mathbb Z}

\def\1{{\bf{1}}}

\usepackage[pdfstartview=FitH]{hyperref}

\def\footnoterule{\kern 1mm \hrule width 7cm \kern 2.2mm}%

\newcommand{\bea}{\begin{eqnarray}}
\newcommand{\eea}{\end{eqnarray}}
\newcommand{\beaa}{\begin{eqnarray*}}
\newcommand{\eeaa}{\end{eqnarray*}}
\newcommand{\be}{\begin{equation}}
\newcommand{\ee}{\end{equation}}
\newcommand{\nn}{\nonumber}

\begin{document}

\title{Symmetric deformed 2D/3D Hurwitz-Kontsevich model and affine Yangian of ${\mathfrak{gl}}(1)$}
\author{Wang Na\dag\footnote{Corresponding author: wangna@henu.edu.cn },\ Wu Ke\ddag\\
\dag\small School of Mathematics and Statistics, Henan University, Kaifeng, 475001, China\\
\ddag\small School of Mathematical Sciences, Capital Normal University, Beijing 100048, China}

\date{}
\maketitle

\begin{abstract}
Since the ($\beta$-deformed) Hurwitz Kontsevich model corresponds to the special case of affine Yangian of ${\mathfrak{gl}}(1)$. In this paper, we construct two general cases of the $\beta$-deformed Hurwitz Kontsevich model. We find that the $W$-operators of these two models can be represented by the generators $e_k,\ f_k,\psi_k$ of the affine Yangian of ${\mathfrak{gl}}(1)$, and the eigenstates (the symmetric functions $Y_\lambda$ and 3-Jack polynomials) can be obtained from the 3D Young diagram representation of affine Yangian of ${\mathfrak{gl}}(1)$. Then we can see that the $W$-operators and eigenstates are symmetric about the permutations of coordinate axes.
\end{abstract}

\section{Introduction}\label{sec1}
The affine Yangian of ${\mathfrak{gl}}(1)$ appears independently in the work
of Maulik-Okounkov \cite{MO} and Schiffmann-Vasserot \cite{SV2} in connection with the AGT conjecture. The affine Yangian of $\mathfrak{gl}(1)$ is isomorphic to the universal enveloping algebra of $W_{1+\infty}$\cite{GGL}.
They can be obtained from the Miura transformation and Yang-Baxter equation\cite{FaLu}. The Miura transformation shows that one can think of $W_N$-algebra as being a quantization of the space of $N$-th order differential operators\cite{Pro1}. The specific free field embedding of $W_N$ in the Fock space of $N$ free bosons depends on the ordering of the Boson fields, we find this ordering corresponds to the ordering $j=1,2,\cdots,N$ of the slices of a 3D Young diagram on the plane $z=j$\cite{WBCW}. The Yang-Baxter equation show that the $R$-matrix $R_{12}$ satisfies\cite{Pro1,Pro}
\[
R_{12}(\alpha_0\partial_z+J_1(z))(\alpha_0\partial_z+J_2(z))=(\alpha_0\partial_z+J_2(z))(\alpha_0\partial_z+J_1(z))R_{12}
\]
where the OPE of Boson field $J_i(z)$ is
\[
J_i(z)J_j(w)\sim \frac{\delta_{i,j}}{(z-w)^2},
\]
The monodromy operator is $\mathcal{T}_A=\mathcal{R}_{A1}\mathcal{R}_{A2}\cdots \mathcal{R}_{AN}$. The Hamiltonian operator $\mathcal{H}$ equals the vacuum expectation of $\mathcal{T}_A$ in the Bosonic Fock space $\mathcal{F}_A$. The eigenstates of $\mathcal{H}$ are 3-Jack polynomials\cite{3-jack}. 

3-Jack polynomials are symmetric functions defined on 3D Young diagrams \cite{3-jack}. In special case $h_1=h,h_2=-h^{-1}$, where $h_1,\ h_2$ are parameters in the affine Yangian of ${\mathfrak{gl}}(1)$, 3-Jack polynomials become Jack polynomials on 2D Young diagrams\cite{3-jack}, and when $h=1$, Jack polynomials become Schur functions\cite{Mac}. Jack polynomials are the eigenstates of the Hurwitz operator $W_0$ in the $\beta$-deformed Hurwitz-Kontsevich model\cite{WLZZ}. The $\beta$-deformed Hurwitz-Kontsevich model, as the deformation of the Kontsevich model (the eigenstates are Schur functions), describes the Hurwitz numbers
and Hodge integrals over the moduli space of complex curves \cite{Shakirov2009,Goulden97,KHurwitz}. $W$-representations give the dual expressions
for the partition functions through differentiation rather than integration \cite{Shakirov2009}. The $W$ operators $W_n, \ n\in\Z$ can be represented by Bosons. The matrix models are generated by the $W$-operators. The Hurwitz-Kontsevich model  (\ref{HKPF}) is a matrix model\cite{Shakirov2009}
\[
Z_{0}\{p\}=\int_{N\times N} \sqrt{{\rm det}\left(\frac{{\rm sinh}(\frac{\phi\otimes I-I\otimes\phi}{2})}
{\frac{\phi\otimes I-I\otimes\phi}{2}} \right)}d\phi e^{-\frac{1}{2t}{\rm Tr}\phi^2
-\frac{N}{2}{\rm Tr}\phi-\frac{1}{6}tN^3+\frac{1}{24}tN+{\rm Tr}(e^{\phi}\psi)},
\]
where  $\psi$ is an $N\times N$ matrix and the time variables $p_k={\rm Tr}\psi^k$, which can be represented 
by the exponent of $W_0$ acting on the function
$e^{p_1/e^{tN}}$.  The
$N\times N$ complex matrix model \cite{AlexandrovJHEP2009,AMironov1705} and the Gaussian hermitian one-matrix model \cite{Shakirov2009,AMironov1705} are generated by $e^{{W}_{-n}/n}\cdot 1$ for $n=1, 2$ respectively. Similarly for the $\beta$-deformed case.

In \cite{HKAYangian}, we find that the $W$ operators in Hurwitz-Kontsevich model can be represented by the generators $e_k,\ f_k, \psi_k$ of the  affine Yangian of $\mathfrak{gl}(1)$ in the special case $h_1=1,h_2=-1$
\be
W_0=\frac{1}{6}\psi_3+\frac{N}{2}\psi_2,\ E_1=e_1+Ne_0,\ E_{-1}=-f_1-Nf_0.
\ee
Then their representation space are the space of Schur functions.
The $W$ operators in $\beta$-deformed Hurwitz-Kontsevich model can be represented by the generators $e_k,\ f_k, \psi_k$ of the  affine Yangian of $\mathfrak{gl}(1)$ in the special case $h_1=h,h_2=-h^{-1}$
\be
{W}_0=\frac{1}{6}\sqrt{\beta}\psi_3+\frac{1}{2}(\beta N-\frac{1}{3}(1-\beta))\psi_2,\ E_1=e_1+\sqrt{\beta} Ne_0,\ E_{-1}=-f_1-\sqrt{\beta}Nf_0,
\ee
where $\beta=h^{-2}$. Then their representation space are the space of Jack polynomials. Other $W$ operators can be obtained from ${W}_0,\ E_1,\ E_{-1}$. 

Clearly, there should be a general Hurwitz-Kontsevich model which corresponds to the  affine Yangian of $\mathfrak{gl}(1)$. In this paper, we construct two general cases of the Hurwitz-Kontsevich model. One is called the symmetric deformed 2D Hurwitz-Kontsevich model, the operators and the eigenstates are symmetric about $x$-axis and $y$-axis in coordinate system $xOy$. The other one is called the symmetric deformed 3D Hurwitz-Kontsevich model, the operators and the eigenstates are symmetric about $x$-axis, $y$-axis and $z$-axis in coordinate system $O-xyz$.  Since the $\beta$-deformed Hurwitz-Kontsevich model describes the Hurwitz numbers
and Hodge integrals over the moduli space of complex curves, the symmetric deformed 2D/3D Hurwitz-Kontsevich model should describe the Hurwitz numbers
and Hodge integrals over a more general case.

 The paper is organized as follows. In section \ref{sect2}, we recall the definition of affine Yangian of ${\mathfrak{gl}}(1)$ from Yang-Baxter equation, which is copied from section 2 of paper \cite{HKAYangian}. In section \ref{sect3}, we construct the symmetric deformed 2D Hurwitz-Kontsevich model, and represent the $W$ operators and eigenstates by the affine Yangian of ${\mathfrak{gl}}(1)$. In section \ref{sect4},  we construct the symmetric deformed 3D Hurwitz-Kontsevich model, and also represent the $W$ operators and eigenstates by the affine Yangian of ${\mathfrak{gl}}(1)$.

\section{Affine Yangian of $\mathfrak{gl}(1)$}\label{sect2}
In this section, we review affine Yangian of $\mathfrak{gl}(1)$ from the Yang-Baxter equation, which is copied from section 2 of \cite{HKAYangian}.
Introduce three complex numbers $h_1,h_2,h_3$ which satisfy $h_1+h_2+h_3=0$, and
\beaa
\sigma_2 &=& h_1 h_2 + h_1 h_3 + h_2 h_3,\\
 \sigma_3 &=& h_1 h_2 h_3.
\eeaa

Let Bosons $a_{j,m}$ satisfy
\be\label{bosonajm}
[a_{i,n},a_{j,m}]=-\frac{h_\sigma}{\sigma_3}\delta_{i,j}m\delta_{n+m,0},
\ee
where $\sigma=1,2,3$. Then the OPE of current $J_j(z)=\sum_{m\in\Z}a_{j,m}z^{-m-1}$ is
\be
J_j(z)J_k(w)\thicksim-\frac{h_\sigma}{\sigma_3}\frac{\delta_{jk}}{(z-w)^2}.
\ee
We say $J_j(z)$ is of type $\sigma$. Remark that here $h_\sigma,\ \sigma=1,2,3$ associate to coordinate axes $x$-axis, $y$-axis, $z$-axis, our results are always symmetric about these three axes, then we always choose $\sigma=3$ to express our results, which means that we choose $z$-axis as the preferred one.

$\mathcal{R}$-matrix satisfies the equation\cite{Pro1}
\be\label{rjj}
\mathcal{R}_{12}(\alpha_0\partial+J_1(z))(\alpha_0\partial+J_2(z))=(\alpha_0\partial+J_2(z))(\alpha_0\partial+J_1(z))\mathcal{R}_{12}.
\ee
where $J_1(z)$ and $J_2(z)$ are of type $\sigma$ and $\tau$ respectively.
Let
\[
 J_{-}(z)=h_\tau J_1(z)-h_\sigma J_2(z)=\sum_{j\in\Z}a_{12,m}z^{-m-1},
\]
where $1$ and $2$ in $a_{12,m}$ correspond to that in $J_1(z)$ and $J_2(z)$. When there is no ambiguity, we omit them.
The OPE of $J_{-}(z)$ is
\be
J_{-}(z)J_{-}(w)\thicksim\rho\frac{\delta_{jk}}{(z-w)^2}
\ee
with \[
\rho=-\frac{h_\tau h_\sigma (h_\tau+h_\sigma)}{\sigma_3}.
\]
Let
\be\label{expensionR}
\mathcal{R}=1+\frac{R^{(1)}}{a_{0}}+\frac{R^{(2)}}{a_{0}^2}+\cdots,
\ee
From (\ref{rjj}), the expression of $R^{(n)}$ can be obtained. One can find that of the first five terms in \cite{Pro1}. We list the first two terms
\bea
R^{(1)}&=&-\sum_{k>0}a_{-k}a_{k},\label{R(1)}\\
R^{(2)}&=&\frac{1}{2}(\sum_{j,k>0}(a_{-j-k}a_{j}a_{k}+a_{-j}a_{-k}a_{j+k}+a_{-j}a_{-k}a_{j}a_{k})+\rho\sum_{j>0}ja_{-j}a_j),\label{R(2)}
\eea
here $a_j$ is $a_{12,j}$ defined above.

The $\mathcal{R}$-matrix satisfies the Yang-baxter equation
\be
\mathcal{R}_{12}\mathcal{R}_{13}\mathcal{R}_{23}=\mathcal{R}_{23}\mathcal{R}_{13}\mathcal{R}_{12}.
\ee
We consider $N$ Fock spaces $\mathcal{F}_j$ and one additional auxiliary space $\mathcal{F}_A$ as in paper \cite{Pro1}. Every current $J_j(z)$ for $j=1,2,\cdots, N$ is of type $\sigma$ which acts on $\mathcal{F}_j$, and $J_A(z)$ is of type $\tau$ which acts on $\mathcal{F}_A$. The monodromy operator
\be\label{monodromy}
\mathcal{T}_A=\mathcal{R}_{A1}\mathcal{R}_{A2}\cdots \mathcal{R}_{AN}
\ee
satisfies
\be\label{YBequation}
\mathcal{R}_{AB}\mathcal{T}_A\mathcal{T}_B=\mathcal{T}_B\mathcal{T}_A\mathcal{R}_{AB},
\ee
which means that $\mathcal{T}_A$ satisfies the Yang-Baxter equation again. Define
\bea
&&\mathcal{H}\equiv\langle 0|_A \mathcal{T}_A |0\rangle_A,\\
&&\mathcal{E}\equiv\langle 0|_A \mathcal{T}_A a_{A,-1}|0\rangle_A,\\ &&\mathcal{F}\equiv\langle 0|_A a_{A,1} \mathcal{T}_A |0\rangle_A,\\
&&\mathcal{H}_\Box\equiv\langle 0|_A a_{A,1} \mathcal{T}_A a_{A,-1}|0\rangle_A.
\eea
The parameters in these currents are $a_{Aj,0}$. We change the parameters to $u$ by the relation
\be
-\frac{a_{Aj,0}\sigma_3}{h_\tau h_\sigma}=u-q_j-\frac{h_\tau-h_\sigma}{2},
\ee
For convenience, we let $q_j=0$, which means that the zero mode $a_{j,0}=0$.

Define
\bea
e(u)&=&h_\tau^{-1}(\mathcal{H}(u))^{-1}\mathcal{E}(u),\\
f(u)&=&-h_\tau^{-1}\mathcal{F}(u)(\mathcal{H}(u))^{-1},\\
\psi (u)&=&h_\tau^{-1}(\mathcal{H}_\Box(u-h_\tau)-\mathcal{E}(u-h_\tau)(\mathcal{H}(u-h_\tau))^{-1}\mathcal{F}(u-h_\tau)))(\mathcal{H}(u-h_\tau))^{-1}.
\eea
They generate affine Yangian of $\mathfrak{gl}(1)$. Introduce the generating functions:
\bea
e(u)=\sum_{j=0}^{\infty} \frac{e_j}{u^{j+1}},\
f(u)=\sum_{j=0}^{\infty} \frac{f_j}{u^{j+1}},\
\psi(u)= 1 + \sigma_3 \sum_{j=0}^{\infty} \frac{\psi_j}{u^{j+1}}.
\eea
The mode operators satisfy\cite{YBalgebra,LV}
\begin{eqnarray}
&&\left[ \psi_j, \psi_k \right] = 0,\\
&&\left[ e_{j+3}, e_k \right] - 3 \left[ e_{j+2}, e_{k+1} \right] + 3\left[ e_{j+1}, e_{k+2} \right] - \left[ e_j, e_{k+3} \right]\nonumber \\
&& \quad + \sigma_2 \left[ e_{j+1}, e_k \right] - \sigma_2 \left[ e_j, e_{k+1} \right] - \sigma_3 \left\{ e_j, e_k \right\} =0,\label{yangian1}\\
&&\left[ f_{j+3}, f_k \right] - 3 \left[ f_{j+2}, f_{k+1} \right] + 3\left[ f_{j+1}, f_{k+2} \right] - \left[ f_j, f_{k+3} \right] \nonumber\\
&& \quad + \sigma_2 \left[ f_{j+1}, f_k \right] - \sigma_2 \left[ f_j, f_{k+1} \right] + \sigma_3 \left\{ f_j, f_k \right\} =0, \label{yangian2}\\
&&\left[ e_j, f_k \right] = \psi_{j+k},\label{yangian3}\\
&& \left[ \psi_{j+3}, e_k \right] - 3 \left[ \psi_{j+2}, e_{k+1} \right] + 3\left[ \psi_{j+1}, e_{k+2} \right] - \left[ \psi_j, e_{k+3} \right]\nonumber \\
&& \quad + \sigma_2 \left[ \psi_{j+1}, e_k \right] - \sigma_2 \left[ \psi_j, e_{k+1} \right] - \sigma_3 \left\{ \psi_j, e_k \right\} =0,\label{yangian4}\\
&& \left[ \psi_{j+3}, f_k \right] - 3 \left[ \psi_{j+2}, f_{k+1} \right] + 3\left[ \psi_{j+1}, f_{k+2} \right] - \left[ \psi_j, f_{k+3} \right] \nonumber\\
&& \quad + \sigma_2 \left[ \psi_{j+1}, f_k \right] - \sigma_2 \left[ \psi_j, f_{k+1} \right] + \sigma_3 \left\{ \psi_j, f_k \right\} =0,\label{yangian5}
\end{eqnarray}

The affine Yangian $\mathcal{Y}$ of ${\mathfrak{gl}}(1)$ is the associative algebra with generators $e_j, f_j$ and $\psi_j$, $j = 0, 1, \ldots$ satisfying the above relations\cite{Pro,Tsy}
and boundary conditions
\begin{eqnarray}
&&\left[ \psi_0, e_j \right]  = 0, \left[ \psi_1, e_j \right] = 0,  \left[ \psi_2, e_j \right]  = 2 e_j ,\label{yangian6}\\
&&\left[ \psi_0, f_j \right]  = 0,  \left[ \psi_1, f_j \right]  = 0,  \left[ \psi_2, f_j \right]  = -2f_j ,\label{yangian7}
\end{eqnarray}
and a generalization of Serre relations
\begin{eqnarray}
&&\mathrm{Sym}_{(j_1,j_2,j_3)} \left[ e_{j_1}, \left[ e_{j_2}, e_{j_3+1} \right] \right]  = 0, \label{yangian8} \\
&&\mathrm{Sym}_{(j_1,j_2,j_3)} \left[ f_{j_1}, \left[ f_{j_2}, f_{j_3+1} \right] \right]  = 0,\label{yangian9}
\end{eqnarray}
where $\mathrm{Sym}$ is the complete symmetrization over all indicated indices which include $6$ terms.

The affine yangian $\mathcal{Y}$ has a representation on the 3D Young diagrams. In order to describe the representation,
as in our paper \cite{3DFermionYangian}, we use the following notations. For a 3D Young diagram $\pi$, the notation $\Box\in \pi^+$ means that this box is not in $\pi$ and can be added to $\pi$. Here ``can be added'' means that when this box is added, it is still a 3D Young diagram. The notation $\Box\in \pi^-$ means that this box is in $\pi$ and can be removed from $\pi$. Here ``can be removed" means that when this box is removed, it is still a 3D Young diagram. For a box $\Box$, we let
\begin{equation}\label{epsilonbox}
h_\Box=h_1y_\Box+h_2x_\Box+h_3z_\Box,
\end{equation}
where $(x_\Box,y_\Box,z_\Box)$ is the coordinate of box $\Box$ in coordinate system $O-xyz$. Here we use the order $y_\Box,x_\Box,z_\Box$ to match that in paper \cite{Pro}.

Following \cite{Pro,Tsy},
introduce
\begin{equation}\label{psi0}
\psi_0(u)=\frac{u+\sigma_3\psi_0}{u}
\end{equation}
and
\be\label{dfnvarphi}
\varphi(u)=\frac{(u+h_1)(u+h_2)(u+h_3)}{(u-h_1)(u-h_2)(u-h_3)}.
\ee
For a 3D Young diagram $\pi$, define $\psi_\pi(u)$ by
\be\label{psipiu}
\psi_\pi(u)=\psi_0(u)\prod_{\Box\in\pi} \varphi(u-h_\Box).
\ee
The representation of affine Yangian on 3D Young diagrams is given by
\bea
\psi(u)|\pi\rangle&=&\psi_\pi(u)|\pi\rangle,\label{psiupi}\\
e(u)|\pi\rangle&=&\sum_{\Box\in \pi^+}\frac{E(\pi\rightarrow\pi+\Box)}{u-h_\Box}|\pi+\Box\rangle,\label{eupi}\\
f(u)|\pi\rangle&=&\sum_{\Box\in \pi^-}\frac{F(\pi\rightarrow\pi-\Box)}{u-h_\Box}|\pi-\Box\rangle\label{fupi}
\eea
where $|\pi\rangle$ means the state characterized by the 3D Young diagram $\pi$ and the coefficients
\be\label{efpi}
E(\pi\rightarrow\pi+\Box)=-F(\pi+\Box\rightarrow\pi)=\sqrt{\frac{1}{\sigma_3} \res_{u \to h_{\Box}} \psi_{\pi}(u)}.
\ee
 Equations (\ref{eupi}) and (\ref{fupi}) mean generators $e_j,\ f_j$ acting on the 3D Young diagram $\pi$ by
\bea
e_j|\pi\rangle &=&\sum_{\Box\in \pi^+}h_\Box^jE(\pi\rightarrow\pi+\Box)|\pi+\Box\rangle,\label{ejpi}\\
f_j|\pi\rangle &=&\sum_{\Box\in \pi^-}h_\Box^jF(\pi\rightarrow\pi-\Box)|\pi-\Box\rangle.\label{fjpi}
\eea

The orthogonality of 3D Young diagrams is $\langle \pi|\pi'\rangle=\delta_{\pi,\pi'}$. In the following, we treat $E(\pi\rightarrow\pi+\Box)|\pi+\Box\rangle$ as one element, still denoted by $|\pi+\Box\rangle$. Then $\langle \pi|\pi\rangle$ does not equal $1$, but it can be calculated by
\bea\label{orthogonality11}
\langle \pi+\Box|\pi+\Box\rangle=E^2(\pi\rightarrow\pi+\Box)\langle \pi|\pi\rangle \ \text{and}\ \langle 0|0\rangle=1.
\eea
The result above means that the symmetric functions associated to 3D Young diagrams are related to the growth processes of 3D Young diagrams. A 3D Young diagram with two different growth processes are linearly related. Remark that the orthogonality (\ref{orthogonality11}) become that of Jack polynomials calculated in \cite{CBWW} in the special case $h_1=h, h_2=-h^{-1}$.

For the use in the following, we calculate the action of $\psi_k$ on 3D Young diagram $|\pi\rangle$
\beaa
\psi_\pi(u)&=&\frac{u+\sigma_3\psi_0}{u}\prod_{\Box\in\pi} \frac{(u-h_\Box+h_1)(u-h_\Box+h_2)(u-h_\Box+h_3)}{(u-h_\Box-h_1)(u-h_\Box-h_2)(u-h_\Box-h_3)}\\
&=&(1+\frac{\sigma_3\psi_0}{u})\prod_{\Box\in\pi}(1+\frac{2\sigma_3}{u^3}+\frac{6h_\Box\sigma_3}{u^4}+o(u^{-4}))
\eeaa
then we obtain
\bea\label{psipi}
\psi_1|\pi\rangle=0,\ \psi_2|\pi\rangle=2|\pi||\pi\rangle,\ \psi_3|\pi\rangle=\sum_{\Box\in\pi}(6h_\Box+2\psi_0\sigma_3)|\pi\rangle,
\eea
where $|\pi|$ is the box number of $\pi$.
Other $\psi_k|\pi\rangle$ can also be calculated this way, we only list this three since we will use them to describe the Hurwitz-Kontsevich model.
\section{The symmetric deformed 2D Hurwitz-Kontsevich model}\label{sect3}
We associate the complex numbers $h_1,\ h_2,\ h_3$ to $y$-axis, $x$-axis, $z$-axis respectively. In this section, we consider 3D Young diagrams which have one layer in $z$-axis direction and treat them as 2D Young diagrams. We want our results are symmetric about $x$-axis and $y$-axis, which means that the results are symmetric about $h_1$ and $h_2$.

Let
$\psi_0=-{1}/{h_1h_2}$ and $p=(p_1,p_2,\cdots)$ with $p_n$ being power sum,
 we have $\langle \pi|\pi\rangle=0$ (defined in (\ref{orthogonality11})) unless 3D Young diagram $\pi$ has one layer in $z$-axis direction, and $N=1$ in (\ref{monodromy}). The eigenstates of Hamiltonian $\mathcal{H}$ are symmetric functions which we denoted by $Y_\lambda(p)$ in \cite{WBCW}. For 2D Young diagrams $(n)$, the symmetric functions $Y_{(n)}=Y_{(n)}(p)$ are determined by
\be\label{y_{(n)}}
\sum_{n\geq 0}\left(\begin{array}{cc} h_2/h_1 \\ n\end{array}\right)Y_{(n)}z^n=\exp(\sum_{n=1}^\infty(-1)^{n-1}\frac{p_n}{n}\frac{h_2}{h_1^n}z^n).
\ee
Define the symmetric operator $\hat{Y}_{(n)}$ by
\be
\sum_{n\geq 0}\left(\begin{array}{cc} h_2/h_1 \\ n\end{array}\right)\hat{Y}_{(n)}z^n=\exp(\sum_{n=1}^\infty(-1)^{n-1}\frac{\text{ad}_{e_1}^{n-1}e_0}{n!}\frac{h_2}{h_1^n}z^n).
\ee
The Pieri formula $Y_{(n)}Y_\lambda$ is defined by
\be\label{pieri}
Y_{(n)}Y_\lambda:=\hat{Y}_{(n)}\cdot Y_\lambda.
\ee
Note that the actions of the generators $e_k,\ f_k,\ \psi_k $ of affine Yangian of ${\mathfrak{gl}}(1)$ are the same with that in (\ref{psiupi})-(\ref{fupi}).

From (\ref{y_{(n)}}) and (\ref{pieri}), we can obtain all the expressions of $Y_\lambda$. For example, from (\ref{y_{(n)}}), we get
\bea
Y_\Box&=&p_1,\\
 Y_{\begin{tikzpicture}
\draw [step=0.2](0,0) grid(.4,.2);
\end{tikzpicture}} &=&\frac{1}{h_1-h_2}p_{2}-\frac{h_2}{h_1-h_2}p_{1}^2,\\
Y_{\begin{tikzpicture}
\draw [step=0.2](0,0) grid(.6,.2);
\end{tikzpicture}}&=&\frac{1}{2h_1-h_2}\frac{1}{h_1-h_2}(2p_{3}-3h_2p_{1}p_{2}+h_2^2p_{1}^3).
\eea
From (\ref{pieri}), we know $Y_\Box Y_\Box=Y_{\begin{tikzpicture}
\draw [step=0.2](0,0) grid(.4,.2);
\end{tikzpicture}}+Y_{\begin{tikzpicture}
\draw [step=0.2](0,0) grid(.2,.4);
\end{tikzpicture}}$, then
\[
Y_{\begin{tikzpicture}
\draw [step=0.2](0,0) grid(.2,.4);
\end{tikzpicture}}=\frac{1}{h_2-h_1}p_{2}-\frac{h_1}{h_2-h_1}p_{1}^2.
\]
Note that the symmetric functions are symmetric about $x$-axis and $y$-axis ($h_1$ and $h_2$), which means that $Y_\lambda$ becomes $Y_{\lambda'}$ when $h_1$ and $h_2$ exchange $h_1\leftrightarrow h_2$, where $\lambda'$ is the conjugate of $\lambda$. For example, the expressions of $ Y_{\begin{tikzpicture}
\draw [step=0.2](0,0) grid(.4,.2);
\end{tikzpicture}}$ and $Y_{\begin{tikzpicture}
\draw [step=0.2](0,0) grid(.2,.4);
\end{tikzpicture}}$ exchange when $h_1$ and $h_2$ exchange.

Since $Y_\lambda$ corresponds to 3D Young diagram $\pi$ with property (\ref{orthogonality11}), then $Y_\lambda$ is dependent on the box growth of $\lambda$. Denote $\begin{tikzpicture}
\draw [step=0.2](0,0) grid(.4,.2);
\draw [step=0.2](0,-0.2) grid(.2,0);
\end{tikzpicture}$ obtained from $\begin{tikzpicture}
\draw [step=0.2](0,0) grid(.4,.2);
\end{tikzpicture}$ by adding one box by $\begin{tikzpicture}
\draw [step=0.2](0,0) grid(.4,.2);
\draw [step=0.2](0,-0.2) grid(.2,0);
\end{tikzpicture}_{h_1h_2}$, while denote $\begin{tikzpicture}
\draw [step=0.2](0,0) grid(.4,.2);
\draw [step=0.2](0,-0.2) grid(.2,0);
\end{tikzpicture}$ obtained from $\begin{tikzpicture}
\draw [step=0.2](0,0) grid(.2,.4);
\end{tikzpicture}$ by adding one box by $\begin{tikzpicture}
\draw [step=0.2](0,0) grid(.4,.2);
\draw [step=0.2](0,-0.2) grid(.2,0);
\end{tikzpicture}_{h_2h_1}$, then we know that
\be
Y_{\begin{tikzpicture}
\draw [step=0.2](0,0) grid(.4,.2);
\draw [step=0.2](0,-0.2) grid(.2,0);
\end{tikzpicture}_{h_1h_2}}=\varphi(h_1-h_2)Y_{\begin{tikzpicture}
\draw [step=0.2](0,0) grid(.4,.2);
\draw [step=0.2](0,-0.2) grid(.2,0);
\end{tikzpicture}_{h_2h_1}}.
\ee
Generally, we have
\be
Y_{\lambda+B+A}(p)=\varphi(h_A-h_B)Y_{\lambda+A+B}(p),
\ee
where $A$ or $B$ denotes a box.

From (\ref{pieri}), we know
\[
Y_\Box Y_{\begin{tikzpicture}
\draw [step=0.2](0,0) grid(.4,.2);
\end{tikzpicture}}=Y_{\begin{tikzpicture}
\draw [step=0.2](0,0) grid(.6,.2);
\end{tikzpicture}}+Y_{\begin{tikzpicture}
\draw [step=0.2](0,0) grid(.4,.2);
\draw [step=0.2](0,-0.2) grid(.2,0);
\end{tikzpicture}_{h_1h_2}},\ \ Y_\Box Y_{\begin{tikzpicture}
\draw [step=0.2](0,0) grid(.2,.4);
\end{tikzpicture}}=Y_{\begin{tikzpicture}
\draw [step=0.2](0,0) grid(.4,.2);
\draw [step=0.2](0,-0.2) grid(.2,0);
\end{tikzpicture}_{h_2h_1}}+Y_{\begin{tikzpicture}
\draw [step=0.2](0,0) grid(.2,.6);
\end{tikzpicture}},
\]
then we have
\beaa
Y_{\begin{tikzpicture}
\draw [step=0.2](0,0) grid(.4,.2);
\draw [step=0.2](0,-0.2) grid(.2,0);
\end{tikzpicture}_{h_1h_2}}
&=&\frac{1}{h_2-2h_1}\frac{1}{h_1-h_2}(2p_{3}-2(h_1+h_2)p_{1}p_{2}+2h_1h_2p_{1}^3),\\
Y_{\begin{tikzpicture}
\draw [step=0.2](0,0) grid(.4,.2);
\draw [step=0.2](0,-0.2) grid(.2,0);
\end{tikzpicture}_{h_2h_1}}
&=&\frac{1}{h_1-2h_2}\frac{1}{h_2-h_1}(2p_{3}-2(h_1+h_2)p_{1}p_{2}+2h_1h_2p_{1}^3),\\
Y_{\begin{tikzpicture}
\draw [step=0.2](0,0) grid(.2,.6);
\end{tikzpicture}}&=&\frac{1}{2h_2-h_1}\frac{1}{h_2-h_1}(2p_{3}-3h_1p_{1}p_{2}+h_1^2p_{1}^3).
\eeaa
Others can be obtained this way.

The power sum operators can be represented by the generators $e_k$ of affine Yangian,
\be
p_n=\frac{1}{(n-1)!}\text{ad}_{e_1}^{n-1}e_0=\frac{1}{n-1}[e_1, p_{n-1}]
\ee
where \[
\text{ad}_{A}^{n}B=\underbrace{[A, [A, \cdots [A }_{n}, B]\cdots ]].
\]
Let
\[
\widetilde{\text{ad}}_A^nB=[\cdots[[ B, \underbrace{A], \cdots A],A ]}_{n},
\]
we have that
\be
\langle p_n,p_n\rangle=\frac{1}{((n-1)!)^2}\langle 0|(-1)^n \widetilde{\text{ad}}_{f_0}^{n-1}f_1 \text{ad}_{e_1}^{n-1}e_0|0\rangle.
\ee
In order to calculate $\langle p_n,p_n\rangle$, we calculate the bracket $[\widetilde{\text{ad}}_{f_0}^{n-1}f_1, \text{ad}_{e_1}^{n-1}e_0]$. When $n=1$,
$[f_0,e_0]=-\psi_0$, we have
\be
\langle p_1,p_1\rangle=\psi_0,\ \text{and}\ \langle p_1^m,p_1^m\rangle=m!\psi_0^m,
\ee
for any $m=1,2,3,\cdots.$
Generally,
\beaa
&&[\widetilde{\text{ad}}_{f_0}^{n-1}f_1, \text{ad}_{e_1}^{n-1}e_0]=[[\widetilde{\text{ad}}_{f_0}^{n-2}f_1,f_1], [e_1,\text{ad}_{e_1}^{n-2}e_0]]\\
&=&-[[f_1,[e_1,\text{ad}_{e_1}^{n-2}e_0]],\widetilde{\text{ad}}_{f_0}^{n-2}f_1]-[[[e_1,\text{ad}_{e_1}^{n-2}e_0],\widetilde{\text{ad}}_{f_0}^{n-2}f_1,f_1],f_1]
\eeaa
the second term equal zero, the first term equals $n(n-1)[\widetilde{\text{ad}}_{f_0}^{n-2}f_1, \text{ad}_{e_1}^{n-2}e_0]$, then we have
\[
\langle p_n,p_n\rangle=\frac{n}{n-1}\langle p_{n-1},p_{n-1}\rangle,
\]
which means
\be\label{pnpnpnmpnm}
\langle p_n,p_n\rangle=n\psi_0,\ \text{and}\ \langle p_n^m,p_n^m\rangle=n^m m!\psi_0^m.
\ee

The conditions in the equation above show that the symmetric functions satisfy
\[
\langle Y_\lambda, Y_\mu\rangle=\delta_{\lambda,\mu} E^2(\phi\rightarrow \Box \rightarrow\cdots\rightarrow\lambda).
\]
We take an example to explain the notations. Let $\lambda=\mu=\begin{tikzpicture}
\draw (0.25,0) -- (0.25,0.25) [-];
\draw (0,0)rectangle(0.5,0.25);
\end{tikzpicture}$,
\beaa
\langle Y_{\begin{tikzpicture}
\draw (0.25,0) -- (0.25,0.25) [-];
\draw (0,0)rectangle(0.5,0.25);
\end{tikzpicture}}, Y_{\begin{tikzpicture}
\draw (0.25,0) -- (0.25,0.25) [-];
\draw (0,0)rectangle(0.5,0.25);
\end{tikzpicture}}\rangle=\langle \frac{1}{h_1-h_2}p_{2}-\frac{h_2}{h_1-h_2}p_{1}^2,\frac{1}{h_1-h_2}p_{2}-\frac{h_2}{h_1-h_2}p_{1}^2\rangle=\frac{2\psi_0}{h_1(h_1-h_2)},
\eeaa
and
\beaa
E^2(\phi\rightarrow \Box\rightarrow \begin{tikzpicture}
\draw (0.25,0) -- (0.25,0.25) [-];
\draw (0,0)rectangle(0.5,0.25);
\end{tikzpicture})=E^2(\phi\rightarrow \Box)E^2(\Box\rightarrow \begin{tikzpicture}
\draw (0.25,0) -- (0.25,0.25) [-];
\draw (0,0)rectangle(0.5,0.25);
\end{tikzpicture})=\psi_0\frac{2}{h_1(h_1-h_2)},
\eeaa
that is,
\[
\langle Y_{\begin{tikzpicture}
\draw (0.25,0) -- (0.25,0.25) [-];
\draw (0,0)rectangle(0.5,0.25);
\end{tikzpicture}}, Y_{\begin{tikzpicture}
\draw (0.25,0) -- (0.25,0.25) [-];
\draw (0,0)rectangle(0.5,0.25);
\end{tikzpicture}}\rangle=E^2(\phi\rightarrow \Box\rightarrow \begin{tikzpicture}
\draw (0.25,0) -- (0.25,0.25) [-];
\draw (0,0)rectangle(0.5,0.25);
\end{tikzpicture}).
\]

From the property (\ref{pnpnpnmpnm}), we have
\be
e^{\sum_{n=1}^\infty\frac{p_k\bar{p}_k}{n\psi_0}}=\sum_{\lambda}\frac{1}{\langle Y_{\lambda},Y_{\lambda}\rangle}Y_{\lambda}\{p\}Y_{\lambda}\{\bar{p}\}.
\ee
Note that the summation of the right hand side is over all Young diagrams, we know that there are many box growth processes of $\lambda$, in the summation we only choose one of them.
Then
\be
e^{-h_1h_2p_1/e^{tN}}=e^{{p_1}/{e^{tN}\psi_0}}=\sum_{\lambda}\frac{1}{\langle Y_{\lambda},Y_{\lambda}\rangle}Y_{\lambda}\{p_k=e^{-tN}\delta_{k,1}\}Y_{\lambda}\{\bar{p}\}.
\ee

The symmetric deformed 2D Hurwitz-Kontsevich model is generated by
\be\label{HKPF}
Z_0\{p\}=e^{tW_0}e^{p_1/\psi_0e^{tN}}=\sum_{\lambda}\frac{e^{tc_\lambda}}{\langle Y_\lambda, Y_\lambda\rangle}Y_\lambda\{p_k=e^{-tN}\delta_{k,1}\}Y_\lambda\{p\}
\ee
with
\[
c_\lambda=\sum_{\Box\in\lambda}(h_{\Box}+\psi_0\sqrt{\beta}N),
\]
where $h_\Box$ is defined in (\ref{epsilonbox}) with $z_\Box=1$ and $t$ is a deformation parameter, and the operator $W_0$ equals
\bea
W_0&=&\frac{1}{2}\sum_{k,l=1}^\infty \left(klp_{k+l}\frac{\partial}{\partial p_{k}}\frac{\partial}{\partial p_{l}}-h_1h_2(k+l)p_kp_l\frac{\partial}{\partial p_{k+l}}\right)\nn\\
&+&\frac{1}{2}\sum_{k=1}^\infty((h_1+h_2)(k-1)+2\psi_0\sqrt{\beta}N) k p_k\frac{\partial}{\partial p_{k}}.
\eea
The symmetric functions $Y_\lambda\{p\}$ will be recalled below. The parameter $\beta$ equals $1/h_1$, or $-h_2$, or their multiplications, such as, $\sqrt{-h_2/h_1}$. We will see that the value of $\beta$ does not affect the eigenstates of $W_0$.

Note that when $h_1=h,h_2=-h^{-1}$, the symmetric deformed 2D Hurwitz-Kontsevich model become the $\beta$-deformed Hurwitz-Kontsevich model in \cite{WLZZ,HKAYangian} different by $\sqrt{\beta}$. When $h=1$, the symmetric deformed 2D Hurwitz-Kontsevich model (\ref{HKPF}) become the Hurwitz-Kontsevich model. 

We define the operator
\begin{equation}
 E_{1}=[W_0, p_1]
=\sum_{n=1}^{\infty}np_{n+1}\frac{\partial}{\partial p_{n}}+ \psi_0\sqrt{\beta} Np_1.
\end{equation}
Then acting on $Y_{\lambda}$
\begin{eqnarray}\label{dE1action}
 E_{1}Y_{\lambda}=\sum_{\Box\in\lambda^+}(h_\Box+\psi_0\sqrt{\beta}N)Y_{\lambda+\Box}.
\end{eqnarray}

The operator ${ W}_{-1}$ is defined to be
${ W}_{-1}=[{W}_0, E_{1}]$.
Acting on symmetric functions $Y_\lambda$,
\bea
{ W}_{-1} Y_\lambda=\sum_{\Box\in\lambda^+}(h_\Box+\psi_0\sqrt{\beta}N)^2Y_{\lambda+\Box}.
\eea

We define a series of operators
\begin{equation}
{W}_{-n}=\frac{1}{(n-1)!}\text{ad}_{W_{-1}}^{n-1} E_1,\ \ n\geq 2.
\end{equation}
When $h_1=h,\ h_2=-h^{-1},\ \sqrt{\beta}=h^{-1}$, operator $W_{-2}$ gives the $W$-operator
in the $W$-representations of $\beta$-deformed Gaussian hermitian matrix model \cite{Morozov1901}.
Similarly to \cite{WLZZ}, the partition function hierarchy with $W$-representations is given by
\begin{equation}
{ Z}_{-n}\{p\}= e^{ W_{-n}/n}\cdot 1, \ \ n\geq 1.
\end{equation}
When  $h_1=h,\ h_2=-h^{-1},\ \sqrt{\beta}=h^{-1}$, it becomes the partition function hierarchy in eq.(48) in \cite{WLZZ}, and specially when $n=1$ and $n=2$, $Z_{-n}\{p\}$ give the $\beta$-deformed rectangular
complex (with $N_1=N_2$) and Gaussian hermitian matrix models \cite{Morozov1901} respectively. These results show that the operators and the symmetric functions in this paper are the generalization of that in \cite{WLZZ}, and have the property that they are symmetric about $x$-axis and $y$-axis. The annihilation operators $E_{-1}$ and $W_n$ can be defined similarly.

In the following of this section, we use the generators of affine Yangian of  ${\mathfrak{gl}}(1)$ to represent the symmetric deformed 2D Hurwitz-Kontsevich model.

 In Yang-Baxter equation (\ref{YBequation}), since $N=1$, we denote the Bosons $a_{1,m}$ by $b_m$, which satisfy
  \be
[b_n,b_m]=\psi_0n\delta_{n+m,0}=-\frac{1}{h_1h_2}n\delta_{n+m,0}.
\ee
 Their actions on symmetric functions $Y_\lambda$ are
\bea
b_{-n}=p_n,\ b_n=-\frac{1}{h_1h_2}n\frac{\partial}{\partial p_n},
\eea
for $n>0$,
and
 \bea
b_{-n}=\frac{1}{(n-1)!}\text{ad}_{e_1}^{n-1}e_0,\ b_{n}=-\frac{1}{(n-1)!}\text{ad}_{f_1}^{n-1}f_0.\label{b_nf_1f_0}
\eea

The generators of affine Yangian can be represented by Boson $b_m$. We list some of them \cite{WBCW}.
\bea
\psi_1&=&0,\\
\psi_2&=&-2h_1h_2\sum_{j>0}b_{-j}b_j=2\sum_{j>0}jp_{j}\frac{\partial}{\partial p_j},\\
\psi_3&=&3h_1^2h_2^2\sum_{j,k>0}(b_{-j-k}b_jb_k+b_{-j}b_{-k}b_{j+k})+3\sigma_3\sum_{j>0}jb_{-j}b_j-\sigma_3\sum_{j>0}b_{-j}b_j,
\eea
and
\bea
e_0= b_{-1},\ e_1=-h_1h_2\sum_{j>0}b_{-j-1}b_j,
\eea
From the relations (\ref{yangian4}) and (\ref{yangian5}), we have
\be
\left[\psi_3,e_k\right]=6e_{k+1}+2\psi_0\sigma_3e_k,\ \left[\psi_3,f_k\right]=-6f_{k+1}-2\psi_0\sigma_3f_k.\label{psi3ek}
\ee
From these results, we describe the symmetric deformed 2D Hurwitz-Kontsevich model by using the generators of affine Yangian. The operator $W_0$ can be expressed by
\be
W_0=\frac{1}{6}\psi_3+\frac{1}{2}(\psi_0\sqrt{\beta} N-\frac{1}{3}\psi_0\sigma_3)\psi_2.
\ee
From (\ref{psipi}),
\beaa
\psi_2Y_\lambda=2|\lambda|Y_\lambda,\ \psi_3Y_\lambda=\sum_{\Box\in\lambda}(6h_\Box+2\psi_0\sigma_3)Y_\lambda,
\eeaa
then
\[
W_0Y_\lambda= \sum_{\Box\in\lambda}(h_{\Box}+\psi_0\sqrt{\beta}N)Y_\lambda,
\]
which matches (\ref{HKPF}).

The operator
\bea
E_1&=&[W_0,p_1]=[\frac{1}{6}\psi_3+\frac{1}{2}(\psi_0\sqrt{\beta} N-\frac{1}{3}\psi_0\sigma_3)\psi_2,e_0]\nn\\
&=& e_1+\psi_0\sqrt{\beta}Ne_0,
\eea
we can check that the action of $E_1$ on $Y_\lambda$ matches that of $e_k$ on $Y_\lambda$. Then all the operators $W_{-n}$ can be represented by $e_k$.

The annihilation operator
\bea
E_{-1}&=&[W_0,\frac{1}{h_1h_2}\frac{\partial}{\partial p_1}]=[\frac{1}{6}\psi_3+\frac{1}{2}(\psi_0\sqrt{\beta} N-\frac{1}{3}\psi_0\sigma_3)\psi_2,f_0]\nn\\
&=&- f_1-\psi_0\sqrt{\beta}Nf_0.
\eea
Then the operator
\beaa
 W_{1}&=&[ W_{0},{ E}_{-1}],\\
 W_n&=& \frac{(-1)^n}{(n-1)!}\text{ad}_{W_1}^{n-1}E_{-1},\ \ n\geq 2,
\eeaa
can be represented by the annihilation operators $f_k$ of affine Yangian of ${\mathfrak{gl}}(1)$.
\section{The symmetric deformed 3D Hurwitz-Kontsevich model}\label{sect4}
In this section, the symmetric deformed 3D Hurwitz-Kontsevich model will be constructed. We will see the symmetric deformed 3D Hurwitz-Kontsevich model (the operators and the eigenstates) are symmetric about $x$-axis, $y$-axis and $z$-axis. This model corresponds to the general case of Yang-Baxter equation (\ref{YBequation}). The number $N$ in monodromy matrix (\ref{monodromy}) means that the 3D Young diagrams have at most $N$ layers in $z$-axis direction. The Yang-Baxter equation (\ref{YBequation}) cut a 3D Young diagram into a series of 2D Young diagrams by plane $z=j$ for $j=1,2,\cdots, N$. Then $z$-axis is a preferred one.

The number $N$ shows
\be
\psi_0=-\frac{N}{h_1h_2}
\ee
in this section.
The power sum variables associated to the 2D Young diagram on the plane $z=j$ are denoted by $p_{j,n}$, which satisfy
\be
\langle p_{j,n},p_{i,m}\rangle=\delta_{j,i}\delta_{n,m}n\psi_0.
\ee
Then the 3-Jack polynomials recalled in Appendix satisfy
\be
\langle \tilde{J}_\pi,\tilde{J}_{\pi'}\rangle=\delta_{\pi,\pi'}E^2(\phi\rightarrow \Box\rightarrow\cdots\rightarrow \pi).
\ee

From Yang-Baxter equation (\ref{YBequation}), the generators of affine Yangian of  $\mathfrak{gl}(1)$ can be represented by the Bosons $a_{j,m}$ defined in (\ref{bosonajm}) with $\sigma=3$.  By replacing $a_{j,m}$ by its representation
\be
a_{j,-m}=p_{j,m},\ \ a_{j,m}=-\frac{1}{h_1h_2}m\frac{\partial}{\partial p_{j,m}},\ \ m\geq 0
\ee
we list some of them \cite{WBCW}
\bea
\psi_1&=&0,\\
\psi_2&=&2\sum_{j=1}^N\sum_{k>0}kp_{j,k}\frac{\partial}{\partial p_{j,k}},\label{psi2bosonpjk}\\
\psi_3 &=&3\sum_{i=1}^N(\sum_{j,k>0}(jkp_{i,j+k}\frac{\partial}{\partial p_{i,j}}\frac{\partial}{\partial p_{i,k}}
-h_1h_2(j+k)p_{i,j}p_{i,k}\frac{\partial}{\partial p_{i,j+k}})\nn\\
&&-6h_3\sum_{i_1<i_2}\sum_{k>0}k^2p_{i_1,k}\frac{\partial}{\partial p_{i_2,k}}-(-4N+6j-3)h_3\sum_{j=1}^N\sum_{k>0}kp_{j,k}\frac{\partial}{\partial p_{j,k}}\nn\\
&&-3h_3\sum_{j=1}^N\sum_{k>0}k^2p_{j,k}\frac{\partial}{\partial p_{j,k}},\label{psi3bosonpjk}
\eea
and
\bea
e_0&=&\sum_{j=1}^N p_{j,1},\\
e_1&=& \sum_{j=1}^N\sum_{k>0}p_{j,k+1}\frac{\partial}{\partial p_{j,k}}.
\eea

Introduce the variables $P_{n,j}$, $n=1,2,\cdots$ and $j=1,2,\cdots,n$. From \cite{YZ}, the variables $P_{n,j}$ are one to one correspondence with 3D Young diagrams. Then 3-Jack polynomials should be polynomials of variables $P_{n,j}$.
Let
\beaa
P_{1,1}&=& p_{1,1}+p_{2,1}+\cdots +p_{N,1},\\
P_{2,1}&=& p_{1,2}+p_{2,2}+\cdots +p_{N,2},\\
P_{2,2}&=& -h_1h_2\sum_{j=1}^N p_{j,1}^2+\frac{h_1h_2}{N}(\sum_{j=1}^N p_{j,1})^2-\sum_{j=1}^N(N-2j+1)h_3p_{j,2}.
\eeaa
Then 3-Jack polynomials in equations (\ref{3jackj11},\ref{3jackj1,1},\ref{3jackj2}) become
\beaa
\tilde{J}_{\scalebox{0.06}{\includegraphics{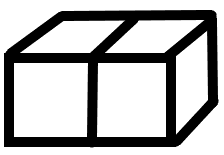}}}&=& \frac{1}{(h_1-h_2)(h_1-h_3)}(\psi_0^{-1}(1+h_2h_3\psi_0)P_{1,1}^2+(1+h_2h_3\psi_0)h_1P_{2,1}+P_{2,2}),\\
 \tilde{J}_{\scalebox{0.08}{\includegraphics{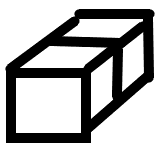}}}&=& \frac{1}{(h_2-h_1)(h_2-h_3)}(\psi_0^{-1}(1+h_1h_3\psi_0)P_{1,1}^2+(1+h_1h_3\psi_0)h_2P_{2,1}+P_{2,2}),\\
 \tilde{J}_{\scalebox{0.06}{\includegraphics{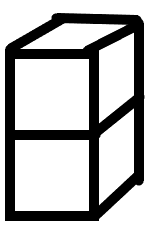}}}&=& \frac{1}{(h_3-h_1)(h_3-h_2)}(\psi_0^{-1}(1+h_1h_2\psi_0)P_{1,1}^2+(1+h_1h_2\psi_0)h_3P_{2,1}+P_{2,2}).
\eeaa
We can see that 3-Jack polynomials are symmetric about $x$-axis, $y$-axis and $z$-axis after changing the variables $\{p_{j,n}\}$ to $\{P_{n,j}\}$.

The general relations between $P_{n,j}$ and $p_{j,n}$ can be obtained from 
\be
P_{n,1}=\frac{1}{(n-1)!}\text{ad}_{e_1}^{n-1}e_0
\ee 
and the orthogonality $\langle P_{n,j},P_{m,i}\rangle=0$ unless $n=m,\ j=i$. Then we have
\be
e^{\sum_{j\leq n}\frac{P_{n,j}\bar{P}_{n,j}}{\langle P_{n,j},P_{n,j}\rangle }}=\sum_{\pi}\frac{1}{\langle \tilde{J}_\pi,\tilde{J}_\pi\rangle}\tilde{J}_\pi\{P_{n,j}\}\tilde{J}_\pi\{\bar{P}_{n,j}\}.
\ee

From the formula above, we obtain
\be
e^{\frac{P_{1,1}}{\psi_0e^{tM}}}=\sum_{\pi}\frac{1}{\langle \tilde{J}_\pi,\tilde{J}_\pi\rangle}\tilde{J}_\pi\{\bar{P}_{n,j}=\delta_{n,1}\delta_{j,1}e^{-tM}\}\tilde{J}_\pi\{{P}_{n,j}\}.
\ee
The symmetric deformed 3D Hurwitz-Kontsevich model is generated by
\begin{equation}
{\mathcal Z}_{0}\{p\}
=e^{t\mathcal{W}_0}\cdot e^{ \frac{P_{1,1}}{\psi_0e^{tM}}},
\end{equation}
where
\begin{eqnarray}
\mathcal{W}_0&=&\frac{1}{2}\sum_{i=1}^N\sum_{k,l=1}^{\infty}\big(klp_{i,k+l}\frac{\partial}{\partial p_{i,k}}\frac{\partial}{\partial p_{i,l}}-h_1h_2(k+l)p_{i,k}p_{i,l}
\frac{\partial}{\partial p_{i,k+l}}\big)\nn\\
&&+(h_1+h_2)\sum_{i_1<i_2}\sum_{k>0}k^2p_{i_1,k}\frac{\partial}{\partial p_{i_2,k}}\nonumber\\
&&+\frac{1}{2}\sum_{j=1}^N\sum_{k=1}^{\infty}\big((h_1+h_2)(k-2N+2j-1)+2\psi_0\sqrt{\beta} N\big)kp_{j,k}\frac{\partial}{\partial p_{j,k}}.\label{w0Jkoperator}
\end{eqnarray}
We can see that when $N=1$, the symmetric deformed 3D Hurwitz-Kontsevich model becomes the symmetric deformed 2D Hurwitz-Kontsevich model, which corresponds to that the 3D Young diagrams which have one layer in $z$-axis direction become 2D Young diagrams.

From 
\be\label{w0jpi}
\mathcal{W}_0\tilde{J}_\pi\{{P}_{n,j}\}=c_\pi \tilde{J}_\pi\{{P}_{n,j}\}
\ee
with \be
c_\pi=\sum_{\Box\in\pi}(h_\Box+\psi_0\sqrt{\beta}N),
\ee
where $h_\Box$ is defined in (\ref{epsilonbox}) and $\psi_0=-N/h_1h_2$,
the partition function ${\mathcal Z}_{0}\{p\}$ equals
\be
\mathcal Z_{0}\{p\}
=\sum_{\pi}\frac{e^{t{c}_\pi}}{\langle \tilde{J}_\pi,\tilde{J}_\pi\rangle}\tilde{J}_\pi\{\bar{P}_{n,j}=\delta_{n,1}\delta_{j,1}e^{-tM}\}\tilde{J}_\pi\{{P}_{n,j}\}.
\ee

Define the creation operator 
\begin{equation}
\bar E_{1}=[\mathcal{W}_0, p_1]
=\sum_{j=1}^N\sum_{k=1}^{\infty}kp_{j,k+1}\frac{\partial}{\partial p_{j,k}}+\psi_0\sqrt{ \beta} N\sum_{j=1}^Np_{j,1}.
\end{equation}
Acting on 3-Jack polynomials,
\be
\bar{E}_1 \tilde{J}_\pi=\sum_{\Box\in\pi^+}(h_\Box+\psi_0\sqrt{\beta}N)\tilde{J}_{\pi+\Box}.
\ee

Define the creation operator
\begin{equation}
{\mathcal W}_{-1}=[\mathcal{W}_0, \bar E_{1}],
\end{equation}
its action on 3-Jack polynomials can be obtained from the actions of $\mathcal{W}_0$ and $\bar E_{1}$ on 3-Jack polynomials
\bea
{\mathcal W}_{-1} \tilde{J}_\pi&=&(\mathcal{W}_0 \bar E_{1}-\bar E_{1}\mathcal{W}_0)\tilde{J}_\pi\nn\\
&=&\sum_{\Box\in\pi^+}(h_\Box+\psi_0\sqrt{\beta}N)\sum_{\Box'\in \pi+\Box}(h_{\Box'}+\psi_0\sqrt{\beta}N)\tilde{J}_{\pi+\Box}\nn\\
&&-\sum_{\Box\in\pi^+}(h_\Box+\psi_0\sqrt{\beta}N)\sum_{\Box'\in \pi}(h_{\Box'}+\psi_0\sqrt{\beta}N)\tilde{J}_{\pi+\Box}\nn\\
&=&\sum_{\Box\in\pi^+}(h_\Box+\psi_0\sqrt{\beta}N)^2\tilde{J}_{\pi+\Box}.
\eea

We define a series of operators
\begin{equation}
{\mathcal W}_{-n}=\frac{1}{(n-1)!}\text{ad}_{\mathcal W_{-1}}^{n-1} \bar{E}_1,\ \ n\geq 2.
\end{equation}
Their actions on 3-Jack polynomials are determined by the actions of $\mathcal{W}_{-1}$ and $\bar E_{1}$ on 3-Jack polynomials.

We introduce the partition function hierarchy with $W$-representations
\begin{equation}
{\mathcal Z}_{-n}\{p\}= e^{\mathcal W_{-n}/n}\cdot 1, \ \ n\geq 1.
\end{equation}
From the discussion in the last section, we know that when $n=1$ and $n=2$, the partition function ${\mathcal Z}_{-n}\{p\}$ is the generalization of the $\beta$-deformed rectangular
complex (with $N_1=N_2$) and Gaussian hermitian matrix models \cite{Morozov1901} respectively to the three dimensional case.

We construct the annihilation operator
\bea
\bar E_{-1}=[\mathcal W_0, -\psi_0\frac{\partial }{\partial P_{1,1}}]=[\mathcal W_0, \frac{1}{h_1h_2}\sum_{j=1}^N\frac{\partial }{\partial p_{j,1}}],
\eea
and 
\begin{equation}
\mathcal W_{1}=[\mathcal W_{0},{\bar E}_{-1}],
\end{equation}
then we introduce a series of the annihilation operators
\begin{equation}
\mathcal{W}_{n}=\frac{(-1)^n}{(n-1)!}\text{ad}_{\mathcal W_{1}}^{n-1} \bar{E}_{-1},\ \ n\geq 2.
\end{equation}

From the results above, we see that the symmetric deformed 3D Hurwitz-Kontsevich model (the operators and the eigenstates) is symmetric about $x$-axis and $y$-axis, but not symmetric about $x$-axis and $z$-axis, or $y$-axis and $z$-axis. In the following, we show that the symmetric deformed 3D Hurwitz-Kontsevich model can be represented by affine Yangian of ${\mathfrak{gl}}(1)$, then we will see that this model is symmetric about $x$-axis, $y$-axis and $z$-axis.

From (\ref{psi2bosonpjk}) and (\ref{psi3bosonpjk}), we have 
\be
\mathcal{W}_0=\frac{1}{6}\psi_3+\frac{1}{2}(\psi_0\sqrt{\beta} N-\frac{1}{3}\psi_0\sigma_3)\psi_2.
\ee
Acting on 3-Jack polynomials,
\be
\mathcal{W}_0\tilde{J}_{\pi}=\frac{1}{6}\psi_3\tilde{J}_\lambda+\frac{1}{2}(\psi_0\sqrt{\beta} N-\frac{1}{3}\psi_0\sigma_3)\psi_2\tilde{J}_\lambda=\sum_{\Box\in\pi}(h_\Box+\psi_0\sqrt{\beta}N)\tilde{J}_\lambda,
\ee
which matches that in (\ref{w0jpi}).

The creation operator
\begin{equation}
\bar E_{1}=[\mathcal{W}_0, p_1]=[\frac{1}{6}\psi_3+\frac{1}{2}(\psi_0\sqrt{\beta} N-\frac{1}{3}\psi_0\sigma_3)\psi_2, e_0]
=e_1+\psi_0\sqrt{\beta}Ne_0.
\end{equation}
From the representation of affine Yangian of ${\mathfrak{gl}}(1)$ on 3D Young diagrams, the action of $\bar E_{1}$ on 3-Jack polynomials is
\be
\bar{E}_1 \tilde{J}_\pi=(e_1+\psi_0\sqrt{\beta}Ne_0) \tilde{J}_\pi=\sum_{\Box\in\pi^+}(h_\Box+\psi_0\sqrt{\beta}N)\tilde{J}_{\pi+\Box}.
\ee

The creation operator
\bea
{\mathcal W}_{-1}&=&[\frac{1}{6}\psi_3+\frac{1}{2}(\psi_0\sqrt{\beta} N-\frac{1}{3}\psi_0\sigma_3)\psi_2, e_1+\psi_0\sqrt{\beta}Ne_0]\nn\\
&=& e_2+2\psi_0\sqrt{\beta}Ne_1+\psi_0\beta N^2e_0,
\eea
then ${\mathcal W}_{-n}, \ {n\geq 1}$ can be represented by the generators $e_k$ of affine Yangian of ${\mathfrak{gl}}(1)$.

The annihilation operator
\begin{equation}
\bar E_{1}=[\frac{1}{6}\psi_3+\frac{1}{2}(\psi_0\sqrt{\beta} N-\frac{1}{3}\psi_0\sigma_3)\psi_2, f_0]
=-f_1-\psi_0\sqrt{\beta}Nf_0.
\end{equation}
Then ${\mathcal W}_{n},\ n\geq 1$ can be represented by the generators $f_k$ of affine Yangian of ${\mathfrak{gl}}(1)$.
\section*{Appendix: 3-Jack polynomials}
3-Jack polynomials are eigenstates of the Hamiltonian $\mathcal{H}$ defined in section \ref{sect2}. 3-Jack polynomials are the generalization of the symmetric functions $Y_\lambda$ to the three dimensional case, which are also the generalization of Jack polynomials. Following the name and the notations we defined in paper \cite{3-jack}, we denote 3-Jack polynomials by $\tilde{J}_\pi$, where $\pi$ is a 3D Young diagram. When $N$ in monodromy matrix (\ref{monodromy}) equals $1$, 3-Jack polynomials become $Y_\lambda$, and when $N=1$ and $h_1=h, h_2=-h^{-1}, \alpha=1/\beta=h^2$, 3-Jack polynomials become Jack polynomials $\tilde{J}_\lambda$. Jack polynomials $\tilde{J}_\lambda$ are defined in \cite{CBWW}, which equal $J_\lambda$ (defined in book \cite{Mac}) multiplied by a constant.

We recall the expressions of 3-Jack polynomials for some of 3D Young diagrams\cite{3-jack}.
\bea
\tilde{J}_{{\scalebox{0.09}{\includegraphics{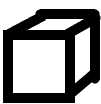}}}}&=&p_{1,1}+\cdots+p_{N,1},\\
\tilde{J}_{\scalebox{0.06}{\includegraphics{Young1.PNG}}}&=& \frac{1}{(h_1-h_2)(h_1-h_3)}(-h_2(h_1-h_3)\sum_{i=1}^Np_{i,1}^2+\sum_{i=1}^N(h_1-(2N-2i+1)h_3)p_{i,2}\nn\\
&&+2h_2h_3\sum_{i_1<i_2}p_{i_1,1}p_{i_2,1} ),\label{3jackj11}\\
 \tilde{J}_{\scalebox{0.08}{\includegraphics{Young2.PNG}}}&=& \frac{1}{(h_2-h_1)(h_2-h_3)}(-h_1(h_2-h_3)\sum_{i=1}^Np_{i,1}^2+\sum_{i=1}^N(h_2-(2N-2i+1)h_3)p_{i,2}\nn\\
&&+2h_1h_3\sum_{i_1<i_2}p_{i_1,1}p_{i_2,1}), \label{3jackj1,1}\\
 \tilde{J}_{\scalebox{0.06}{\includegraphics{Young3.PNG}}}&=& \frac{1}{(h_3-h_1)(h_3-h_2)}(-2\sum_{i=1}^{N-1}(N-i)h_3p_{i,2}+2h_1h_2\sum_{i_1<i_2}p_{i_1,1}p_{i_2,1}),\label{3jackj2}
\eea
and 
\bea
\tilde{J}_{\scalebox{0.07}{\includegraphics{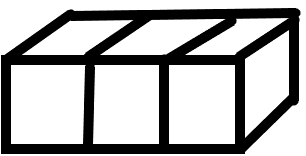}}}&=& \frac{1}{(h_1-h_2)(h_1-h_3)(2h_1-h_2)(2h_1-h_3)}(h_2^2(6h_1^2+5h_1h_2+h_2^2)\sum_{j=1}^N p_{j,1}^3,\nn\\
&&-3h_2(2h_1+h_2)\sum_{j=1}^N((2(N-j)+3)h_1+(2(N-j)+1)h_2)p_{j,1}p_{j,2}\nn\\
&&+2\sum_{j=1}^N(3(N-j+1)(N-j+2)h_1^2+(6(N-j)^2+12(N-j)+5)h_1h_2\nn\\
&&+(3(N-j)(N-j+1)+1)h_2^2)p_{j,3}-3h_2^2h_3(2h_1+h_2)\sum_{i<j}(p_{i,1}^2p_{j,1}+p_{i,1}p_{j,1}^2)\nn\\
&&+3h_2h_3\sum_{i<j}((2N-2j+2)h_1+(2N-2j+1)h_2)p_{i,1}p_{j,2}\nn\\
&&+3h_2h_3\sum_{i<j}((2N-2i+4)h_1+(2N-2i+1)h_2)p_{i,2}p_{j,1}\nn\\
&&+6h_2^2h_3^2\sum_{j_1<j_2<j_3}p_{j_1,1}p_{j_2,1}p_{j_3,1})
\eea
\bea
\tilde{J}_{{\scalebox{0.08}{\includegraphics{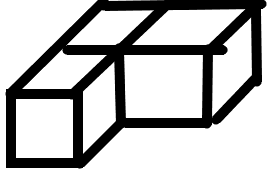}}}_{h_1,h_2}}&=& \frac{1}{(h_1-h_2)(h_1-h_3)(h_2-2h_1)(h_2-h_3)}(2h_1h_2(2h_1^2+5h_1h_2+2h_2^2)\sum_{j=1}^N p_{j,1}^3\nn\\
&&+2h_3\sum_{j=1}^N((2N-2j+2)h_1^2+(7N-7j+5)h_1h_2+(2N-2j+2)h_2^2)p_{j,1}p_{j,2}\nn\\
&&+2\sum_{j=1}^N((3(N-j)^2+5(N-j)+2)h_1^2+(6(N-j)^2+10(N-j)+5)h_1h_2\nn\\
&&+(3(N-j)^2+5(N-j)+2)h_2^2)p_{j,3}+8h_1h_2h_3^2\sum_{i<j}(p_{i,1}^2p_{j,1}+p_{i,1}p_{j,1}^2)\nn\\
&&+2h_3\sum_{i<j}((2N-2j+2)h_1^2+(4N-4j+1)h_1h_2+(2N-2j+2)h_2^2)p_{i,1}p_{j,2}\nn\\
&&+4h_3^3\sum_{i<j}(N-i+1)p_{i,2}p_{j,1}+12h_1h_2h_3^2\sum_{j_1<j_2<j_3}p_{j_1,1}p_{j_2,1}p_{j_3,1})
\eea
\bea
 \tilde{J}_{{\scalebox{0.08}{\includegraphics{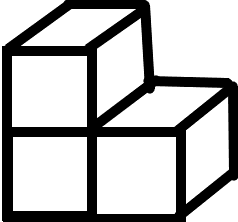}}}_{h_1,h_3}}&=& \frac{1}{(h_1-h_2)(h_1-h_3)(h_3-2h_1)(h_3-h_2)}(2h_2h_3(3h_1+2h_2)\sum_{j=1}^N (N-j)p_{j,1}p_{j,2}\nn\\
&&-2h_3\sum_{j=1}^N(3(N-j)(N-j+1)h_1+(N-j)(3N-3j+1)h_2)p_{j,3}\nn\\
&&-2h_1h_2^2(3h_1+2h_2)\sum_{i<j}(p_{i,1}^2p_{j,1}+p_{i,1}p_{j,1}^2)\nn\\
&&+2h_2\sum_{i<j}(3h_1^2-2(N-j-1)h_1h_2-(2N-2j)h_2^2)p_{i,1}p_{j,2}\nn\\
&&-2h_2^2\sum_{i<j}((2N-2i+1)h_1+(2N-2i)h_2)p_{i,2}p_{j,1}\nn\\
&&+12h_1h_3h_2^2\sum_{j_1<j_2<j_3}p_{j_1,-1}p_{j_2,-1}p_{j_3,-1})
 \eea
 \bea
  \tilde{J}_{{\scalebox{0.09}{\includegraphics{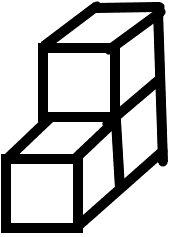}}}_{h_2,h_3}}&=& \frac{1}{(h_2-h_1)(h_2-h_3)(h_3-2h_2)(h_3-h_1)}(2h_1h_3(3h_2+2h_1)\sum_{j=1}^N (N-j)p_{j,1}p_{j,2}\nn\\
&&-2h_3\sum_{j=1}^N(3(N-j)(N-j+1)h_2+(N-j)(3N-3j+1)h_1)p_{j,3}\nn\\
&&-2h_2h_1^2(3h_2+2h_1)\sum_{i<j}(p_{i,1}^2p_{j,1}+p_{i,1}p_{j,1}^2)\nn\\
&&+2h_1\sum_{i<j}(3h_2^2-2(N-j-1)h_1h_2-(2N-2j)h_1^2)p_{i,1}p_{j,2}\nn\\
&&-2h_1^2\sum_{i<j}((2N-2i+1)h_2+(2N-2i)h_1)p_{i,2}p_{j,1}\nn\\
&&+12h_1h_3h_1^2\sum_{j_1<j_2<j_3}p_{j_1,1}p_{j_2,1}p_{j_3,1})
  \eea
  \bea
  \tilde{J}_{\scalebox{0.08}{\includegraphics{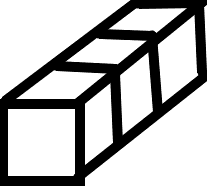}}}&=& \frac{1}{(h_2-h_1)(h_2-h_3)(2h_2-h_1)(2h_2-h_3)}(h_1^2(6h_2^2+5h_1h_2+h_1^2)\sum_{j=1}^N p_{j,1}^3\nn\\
&&-2h_1(2h_2+h_1)\sum_{j=1}^N((2(N-j)+1)h_2+(2(N-j)+1)h_1)p_{j,1}p_{j,2}\nn\\
&&+2\sum_{j=1}^N(3(N-j+1)(N-j+2)h_2^2+(6(N-j)^2+12(N-j)+5)h_1h_2\nn\\
&&+(3(N-j)(N-j+1)+1)h_1^2)p_{j,3}-3h_1^2h_3(2h_2+h_1)\sum_{i<j}(p_{i,1}^2p_{j,1}+p_{i,1}p_{j,1}^2)\nn\\
&&+3h_1h_3\sum_{i<j}((2N-2j+2)h_2+(2N-2j+1)h_1)p_{i,1}p_{j,2}\nn\\
&&+3h_1h_3\sum_{i<j}((2N-2i+4)h_2+(2N-2i+1)h_1)p_{i,2}p_{j,1}\nn\\
&&+6h_1^2h_3^2\sum_{j_1<j_2<j_3}p_{j_1,1}p_{j_2,1}p_{j_3,1})
\eea
 \bea
  \tilde{J}_{\scalebox{0.08}{\includegraphics{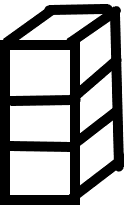}}}&=& \frac{1}{(h_3-h_1)(h_3-h_2)(2h_3-h_1)(2h_3-h_2)}(
6h_3^2\sum_{j=1}^N(N-j)(N-j-1)p_{j,3}\nn\\
&&-6h_1h_2h_3\sum_{i<j}(N-j)p_{i,1}p_{j,2}-6h_1h_2h_3\sum_{i<j}(N-i-1)p_{i,2}p_{j,1}\nn\\
&&+6h_1^2h_2^2\sum_{j_1<j_2<j_3}p_{j_1,1}p_{j_2,1}p_{j_3,1}).
\eea

 \section*{Data availability statement}
The data that support the findings of this study are available from the corresponding author upon reasonable request.

\section*{Declaration of interest statement}
The authors declare that we have no known competing financial interests or personal relationships that could have appeared to influence the work reported in this paper.

\section*{Acknowledgements}
This research is supported by the National Natural Science Foundation
of China under Grant No. 12101184 and No. 11871350, and supported by Key Scientific Research Project in Colleges and Universities of Henan Province No. 22B110003.

\end{document}